\documentclass[12pt,preprint]{aastex}

\shorttitle{Evidence that the Sun was Born in a Cluster}
\shortauthors{Looney et al.}

\newcommand{\msun}{\mbox{$M_{\sun}$}}
\def\iso#1#2{\mbox{${}^{#2}{\rm #1}$}}
\def\fe#1{\iso{Fe}{#1}}
\def\finj{f_{\rm inj}}

\def\pfrac#1#2{\left( \frac{#1}{#2} \right)}

\begin{document}

\title{Radioactive Probes of the Supernova-Contaminated Solar Nebula:
Evidence that the Sun was Born in a Cluster}
\author{Leslie W. Looney, John J. Tobin, and Brian D. Fields}
\affil{Department of Astronomy, University of Illinois at Urbana-Champaign, IL 61801}

\begin{abstract}
We 
construct a simple model for radioisotopic enrichment of the protosolar nebula
by injection from a nearby supernova,
based on the inverse square law for ejecta dispersion.
This idealized model concisely and explicitly 
connects the observational data to the
key astrophysical parameters; as such, 
it provides a useful tool for parameter studies that
can inform the computationally
expensive numerical (magneto)hydrodynamic studies essential for
a full solution to this problem.
We find that the presolar radioisotopes
abundances (i.e., in solar masses)
demand a nearby supernova: its distance $D$ can be no larger than
$D/R_{\rm SS} \le 66$ times the size $R_{\rm SS}$
of the protosolar nebula, at a 90\% confidence level, assuming 1 M$_\odot$ of protosolar material.
The relevant size of the nebula depends on its
state of evolution at the time of radioactivity injection.
In one 
scenario, a collection of
low-mass stars, including our sun, formed in a group or cluster with
an intermediate- to high-mass star that ended its life as a supernova
while our sun was still a protostar, a starless core, or perhaps a diffuse cloud. 
Using recent observations of protostars
to estimate the size of the protosolar nebula 
constrains the distance of the supernova
to $D \sim 0.02$ pc to 1.6 pc.
The supernova distance limit
is consistent with the scales of
low-mass stars formation around one or more
massive stars,
but it is closer than expected were the sun formed in an
isolated, solitary state.
Consequently, if {\em any} presolar radioactivities
originated via supernova injection,
we must conclude that our sun was 
a member of such a group or cluster that has 
since dispersed, and thus that solar system formation should
be understood in this context.
In addition, we show that the timescale from explosion to the creation
of small bodies was on the order of 1.8 Myr (formal 90\% confidence range
of 0 to 2.2 Myr), and thus
the temporal choreography from supernova
ejecta to meteorites is important. 
Finally, we can not distinguish between
progenitor masses from 15 to 25 M$_\odot$ in the nucleosynthesis models; 
however, the 20 M$_\odot$ model is somewhat preferred.
\end{abstract}


\section{Introduction}

The picture of star formation that has been developed over the last few
decades provides a good broad-brush explanation of isolated, low-mass,
Sun-like star formation \citep[e.g.][]{shuppiii}.  
However, is that picture valid for the formation of the Sun?
The scenario is probably unlikely as 
there is good evidence that the majority of Sun-like stars
form in clusters \citep[e.g.,][]{carpenter2000,ladalada2003}.
In addition, the theoretical view that protostars form in relative isolation from
their molecular cloud neighbors is also weakening \citep[e.g.,][]{clark2005}.

To better explain the origin of the Sun and Planets, it is important 
to first understand the
initial conditions in which the Sun formed.  Did the Sun form in isolation
or in a small group or cluster of stars that has since dispersed?
The concept of a lonely Solar System birth has been
challenged by the idea that the Sun was
born near at least one massive star, in a group or cluster of stars, based on the study
of short-lived radioisotopes in meteorites 
\citep[e.g.][]{hester2004}.

The main evidence for short-lived radioactive species is from
isotopic anomalies of daughter species in primitive meteorites
\citep[for recent reviews, see][]{goswami2000,meyer2000,wadhwa2006}.
The isotopes are now extinct, but they must have been live,
and in relatively high abundances,  during the
early stages of planet formation in our Solar System.
Possible origins for the measured isotopic abundances
are that the Sun formed in a location that was enriched
either by a stellar wind or a recent core-collapse supernova \citep[e.g.][]{lee1977},
hence in or near a cluster or stellar group environment.
Other possibilities for the short-lived radioactive isotopes
include irradiation in the circumstellar disk
\citep[e.g.,][]{lee1998} or mixing of ejecta from
core-collapse supernova in the interstellar medium (ISM) 
\citep[e.g.,][]{meyer2000,hester2005,oue2005}.  
However, new results reveal that the early Solar System
contained significant amounts of the short-lived $^{60}$Fe radionuclide
\citep[e.g.][]{tachibana2003,most2005} that
can not be explained with the irradiation models or
ISM mixing models \citep[e.g.][]{tachibana2003,gounelle2006}.

In this paper, we make the assumption that a supernova event occurred
near the early solar nebula, such that short-lived radioactive species were injected
into the solar nebula \citep[e.g.][]{cameron1977,goswami2000}.
Although, the $^{60}$Fe abundance could also be explained by ejecta from
Asymptotic Giant Branch (AGB) stars, novae, or Type I supernovae,
the likelihood of such late stellar stage objects overlapping with
a star formation region is much smaller than a supernova that was
born in the same cluster \citep[e.g.][]{kastner1994,hester2005}.

With the assumption of a nearby supernova, 
we construct a simple model that directly and explicitly 
relates the observed pre-solar radioisotope abundances
to the supernova distance and the size of the protosolar
nebula into which they were injected.
Our model, while idealized, clearly illustrates
that the radioisotope abundances directly
constrain (and in particular, place an upper limit on)
the ratio of the supernova distance to the size of 
the nebula.
This type of model has been discussed before 
\citep[e.g.][]{vanhala2000,vanhala2002,oue2005}
with more emphasis on the detailed hydrodynamics of
the injection of supernova material into the nebula.
In this paper, we focus on a statistical parameter study,
with emphasis on the supernova distance $D$.
Our simple model allows for a rapid search of
the parameter space and 
shows that in most scenarios, a nearby supernova
(i.e., $D \le$  a few pc, and $\ll$ 1 kpc)
is required if {\em any} radioisotopes came
from a supernova.
This has important qualitative implications for the nature of
the Sun's formation, and quantitatively 
our results can inform and calibrate the more 
computationally expensive hydrodynamics calculations
needed to solve the problem in all of its detail.

\section{Supernova Radioisotope Injection into the Protosolar Nebula}

The abundance of radioisotopes injected by a supernova
explosion depends on--and thus probes--the properties
of the Type II core-collapse supernova and its explosion, the propagation of
the supernova remnant and entrained ejecta, and the
ability of the protosolar nebula to capture the ejecta
that impinge on it.
Each of these aspects of the problem are known to be
quite complex in its details \citep[e.g.][]{chevalier2000,meyer2000,
vanhala2000,vanhala2002}, and a
full understanding of the problem must embrace this
richness.  Nonetheless, a simplified analytic treatment
of the parameter space of the problem
not only offers physical insight, but also provides a 
benchmark for comparison with the more elaborate descriptions.

\subsection{Radioactivity Distance}

In this spirit, we adopt the following idealized
model \citep[also see,][]{vanhala2000,oue2005}.
We assume a supernova event occurred at a distance $D$ from the early
solar nebula.
The ejecta from the supernova blast is then spread
with a ``flux'' (in fact, surface density) that
follows the inverse square law.
It then follows that 
the mass fraction $X_{i} = M_{{\rm SS},i}/M_{\rm SS}$, 
of a radionuclide $i$, observed in the Solar
System is
\begin{equation}
X_{i} = \finj
\frac{M_{{\rm SN},i}}{M_{\rm SS}} \frac{\pi R_{\rm SS}^2}{4\pi D^2} e^{-t/\tau_i}
\label{gen}
\end{equation}
where 
$M_{{\rm SS},i}$ is the mass of radionuclide $i$ in
the early solar system,
$M_{\rm SS}$ is the effective ``target'' or ``absorbing''
mass of the solar nebula,
$M_{{\rm SN},i}$ is the amount of radionuclide $i$ produced in the supernova,
and $R_{\rm SS}$ is the solar system radius.
The exponential term accounts for the decay of the radionuclide, with
decay constant $\tau_i$, in the time
interval $t$ between the supernova explosion and solar system
formation (more precisely, until meteorite formation).
Finally, the $\finj \le 1$ measures the ``injection efficiency,'' 
i.e., the ratio of incident supernova radioisotope debris that is actually
incorporated into the nebula.
For example, 
\cite{vanhala2000} discuss the injection
process in detail suggesting that in some cases the injection occurs via
Rayleigh-Taylor instabilities with an
efficiency of approximately $f \simeq 0.1$.

This model is simplified, and thus idealized, by
assuming not only spherical, homogenous ejecta from isotropical supernova ejecta, but also that 
the supernova ejecta will be completely integrated and 
well mixed in the solar system material.
On the other hand, the model readily provides insight into 
the proximity of the
supernova that ejected the short-lived isotopes into the early
solar nebula.

From Equation \ref{gen}, we can derive a measure
of ``radioactivity distance'', which
is the ratio of the supernova distance to the solar nebula radius.\footnote{
The radioactivity distance is a close analog of a luminosity distance:
radioisotope nucleosynthetic 
yield $M_{\rm SN,i}$ takes the role of luminosity, 
and solar abundance (in fact, surface density $M_{\rm SS,i}/\pi R_{\rm SS}^2$) 
takes the role of observed flux.
}
\begin{equation}
\frac{D}{R_{\rm SS}} = \frac{1}{2} 
  \sqrt{\finj \frac{M_{\rm SN,i}}{X_{i} M_{\rm SS}}}
  e^{-t/2\tau_i} 
 = 
  100 \ \finj^{1/2} \ e^{-t/2\tau_i}  \
   \pfrac{M_{{\rm SN},i}}{10^{-4} \msun}^{1/2}
   \pfrac{10^{-9}}{X_{i}}^{1/2}
   \pfrac{1 \msun}{M_{\rm SS}}^{1/2}
\label{dist}
\end{equation}
where the fiducial values are appropriate for \fe{60} and 1 M$_\odot$ of solar nebula material.
In other words, the radioactivity distance is only dependent
on four variables.  Arguably, the two with the largest uncertainties
are the supernova yields $M_{\rm SN,i}$ and 
presolar radioisotope abundance $X_{i}$.
The supernova yields $M_{{\rm SN},i}$ 
are gathered from models of nucleosynthesis
in massive stars
\citep[e.g.,][]{woosley95,rauscher02}, 
which depend chiefly on the mass of the progenitor star.
It is interesting to note that
the input parameters
appear under a square root
in the expression for radioactivity distance,
which is correspondingly less sensitive to 
errors in the inputs.

The abundance at isotopic closure of a once-live radionuclide parent ${}^{i}{\cal P}$ 
is obtained from meteoritic measurement of anomalous abundances in
its daughter product ${}^{i}{\cal D}$, where the decay
scheme is ${}^{i}{\cal P} \rightarrow {}^{i}{\cal D}$
\citep[see recent review,][]{wadhwa2006}.
This procedure yields the initial ratio ${}^{i}{\cal P}/{\cal P}_{\rm ref}$
of the radioactive parent nuclide to a stable (and abundant)
reference isotope  ${\cal P}_{\rm ref}$ of the same atomic species.
For example, studies of \iso{Ni}{60} excesses in meteorites
give a measure of the pre-solar \fe{60}/\fe{56} ratio
\citep[e.g.][]{tachibana2003,most2005,tachibana2006}.

Table \ref{data} lists the short-lived radionuclides 
in the early solar nebula that are most likely to be contaminated 
by a Type II supernova ejecta \citep[e.g.][]{wadhwa2006}.
We have excluded $^{92}$Nb and $^{244}$Pu since the former
has very large uncertainties and the latter is not predicted in
the two models used in our analysis.
The ratios used are for the initial solar system abundance, which
is defined at the creation of the calcium-aluminum-rich inclusions (CAIs), not necessarily at isotopic closure.
This, however, creates difficulties when referencing the age at
isotopic closure to the initial solar system abundance for some radionuclides.
The $^{36}$Cl/$^{35}$Cl and $^{41}$Ca/$^{40}$Ca ratios are impacted the most from this uncertainty
and may be considered lower limits \citep[e.g.,][]{wadhwa2006}.

To recover the observed
mass fraction
for the presolar nebula (Table \ref{data}), we use the expression \begin{equation} X_i =
\frac{{}^{i}{\cal P}}{{\cal P}_{\rm ref}} \frac{{\cal P}_{\rm ref}}{\rm
H} A_i X_{\rm H} \end{equation} where $A_i$ is the atomic weight
of the isotope.  The early solar ${\cal P}_{\rm ref}/{\rm H}$ abundance
and the hydrogen mass fraction $X_{\rm H} = 0.711$ were taken from
\cite{lodders2003}, Table 2, where the associated uncertainties from
the meteoritic ratios and abundances have been propagated.
In most cases, the largest uncertainty is from the meteoritic ratios;
the exception of which is the radionuclide ratio $^{129}$I/$^{127}$I, which
has a large abundance uncertainty.

For the supernova radioisotope yields $M_{{\rm SN},i}$, we use
the results of theoretical models.  
To illustrate the (significant) uncertainties in these
difficult calculations, we will compute results using both
the model of \cite{woosley95} (at solar metallicity), which spans the mass
progenitor range $11-40 \msun$, and that of \cite{rauscher02},
which updates much of the physics of the earlier model, 
but spans a narrower mass range.
The model uncertainties are clearly large:
for a given progenitor mass, 
the yields for the same isotope in the two models can
diverge in some cases by as much as a factor of ten.
We thus believe it is not overly conservative to
adopt an uncertainty on the model yields
as $\sigma_{M_{\rm SN,i}} = M_{\rm SN,i}$, i.e., a factor of two.

In addition, although a supernova can
probably occur with stellar masses of 8 to 10 $M_\odot$, and indeed the
lower mass supernova may be the most likely for the case of the Sun,
the fate of these stars, and their nucleosynthesis, remains unclear.
The main reason is that the lower mass objects are likely to behave
differently than the standard core collapse SNe with more associated
{\it r}-processed material \citep{wheeler1998}.  
Unfortunately, these systems are not
well studied, so not further discussed in this paper.
Only the mass range of $11-25 \msun$ is considered, 
which provides results for low-mass supernovae and 
an overlap of the two nucleosynthesis models.  The 30, 35, and 40
$\msun$ models in \cite{woosley95} provide flat yields in
comparison to the 25 $\msun$ model, with the exception of
$^{53}$Mn that varies significantly depending on the
assumed kinetic energy.

\subsection{Results}
\label{sect:multisource}

In this section, we present numerical results for the
radioactivity distance ratio $D/R_{\rm SS}$.
We will assume a perfect injection efficiency $\finj = 1$ and a solar system
nebula mass $M_{SS} = 1 M_\odot$,
in both the text and figures, and we will revisit these assumptions in our
discussion.  Note that the results can be scaled for other 
efficiencies or masses via
$D/R_{\rm SS} \propto \finj^{1/2}$ or $D/R_{\rm SS} \propto {M_{SS}}^{-1/2}$.
In \S \ref{numbers}, these results will be applied to two basic cases:
(1) injection into a core or circumstellar envelope, where $\finj \simeq 0.1$, $M_{SS} \simeq 1 M_\odot$,
and $R_{\rm SS} \simeq 0.05$ pc; and
(2) injection into a circumstellar disk,
where $\finj \simeq 1$, $M_{SS} = M_{disk} \simeq 0.01 M_\odot$,
and $R_{\rm SS} \simeq 100$ AU.

\begin{figure}
\includegraphics{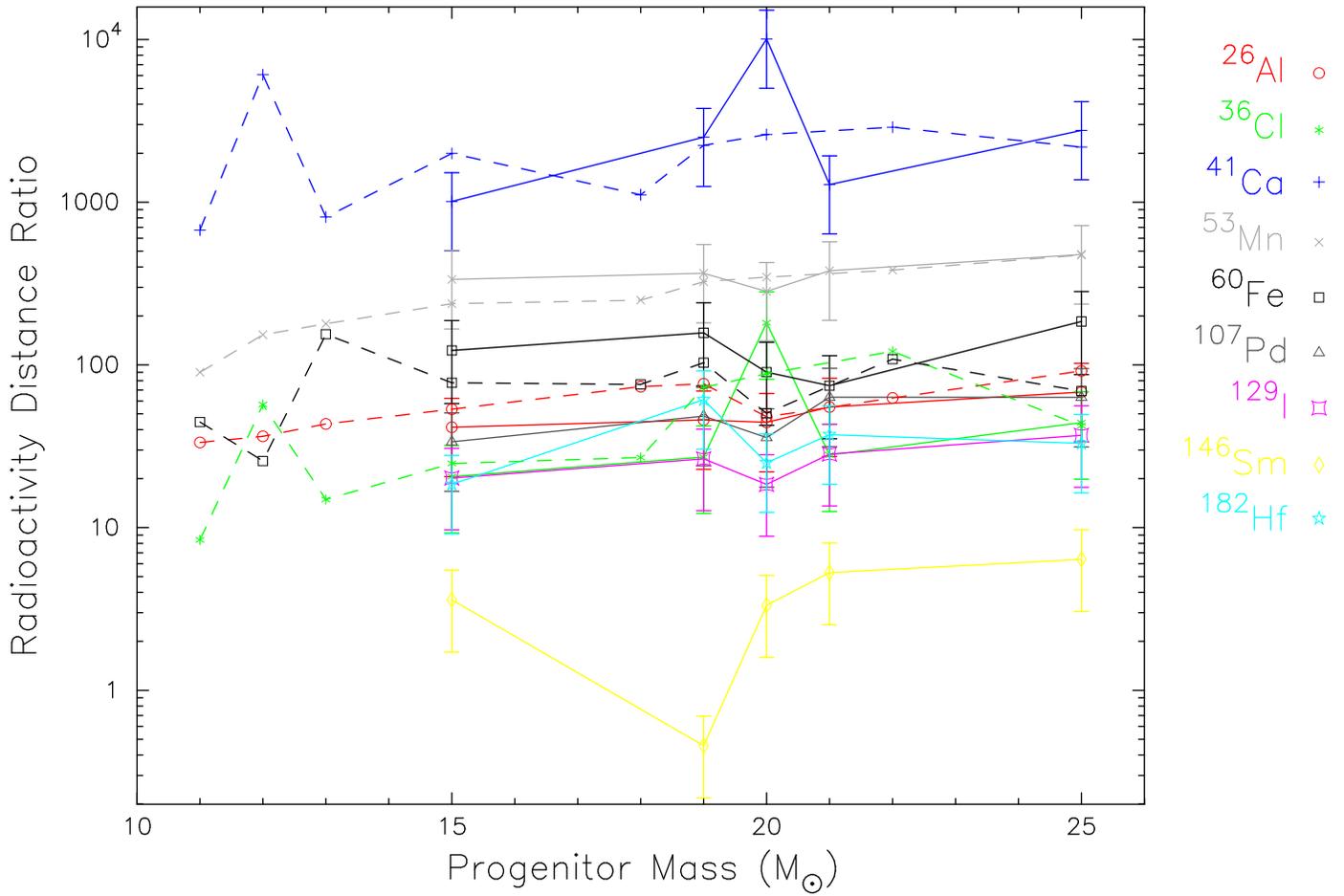}
\vspace{8cm}
\caption{
The radioactive distance ratio without any decay time.  
The supernova production models of \cite{woosley95}, dotted lines, and 
\cite{rauscher02}, solid lines, are used.  The estimated error for each
data point is shown only for the \cite{rauscher02} models in order
to minimize confusion in the Figure.
}
\label{rauscher}
\end{figure}

Figure \ref{rauscher} shows the radioactivity distance for the nine
short-lived radioactive isotopes tabulated in Table \ref{data},
for the case $t/\tau_i \ll 1$ of negligible delay between radioisotope nucleosynthesis
and incorporation into meteorites.
Both nucleosynthesis models are shown: the \cite{rauscher02} model with solid lines and 
the \cite{woosley95} model with dashed lines.  
As the figure is already complicated, the error bars on the \cite{woosley95} model
are not drawn.
Broadly speaking, the order of magnitude of the 
results are in line with the fiducial estimate in Equation~(\ref{dist}),
spanning $D/R_{\rm SS} \sim 10^1 - 10^3$.
We immediately see that if {\em any} pre-solar radioisotopes 
originated in a supernova, 
then (1) the explosion 
had to be close, within 10-10000 solar nebula radii;
and (2) the supernova had to be recent, $t/\tau_i \la 1$, else
many of the shortest lived radioisotopes would have decayed and/or the required distance
would become unfeasibly close.

In more detail, 
the  large range of radioactivity distance in
Figure \ref{rauscher} 
illustrates the complexity of the problem and the
uncertainties in such a model.
The nucleosynthesis uncertainties alone are clearly
very large, as the two supernova models lead to distance estimates
that can vary by factors as high as $\sim 3$.
In the ideal case, in which all radioisotopes did originate
in a supernova as described in the model, 
one would expect the curves for all isotopes
to converge to a single unique value at some progenitor mass;
this is not the case.
In fact, it is clear from Figure \ref{rauscher} that the mass of
the progenitor is unconstrained.
Nonetheless, the majority of the isotopes are clustered near the middle
of the graph, with some isotopes
preferring larger radioactive distances ($^{41}$Ca and $^{53}$Mn) and smaller
radioactive distances ($^{146}$Sm).

Two possible explanations for the stratification in Figure \ref{rauscher}
suggest themselves: multiple sources of short-lived radioisotopes and decay
of the radioisotopes.
We have thus far considered the case of a supernova as the
unique source of presolar radioactivities,
but other sources certainly are possible and indeed
must be present at some level.
One source is the abundance of radioisotopes
in the general interstellar medium,
which achieves a steady state between production
and decay.
The steady-state abundance is essentially
the production abundance, reduced by
the ratio $\tau_i/\tau_\star$
of decay time to a star-formation time $\tau_\star$
appropriate for the global ISM.
Consequently,
the highest steady-state abundances
will be those of long-lived species, such as
$^{53}$Mn,$^{146}$Sm,$^{107}$Pd, and
$^{129}$I and others \citep{meyer2000}.

Another more promising source of presolar radioisotopes
is {\em in situ} production through
protostellar cosmic rays in the circumstellar disk \citep[$^{41}$Ca
and $^{53}$Mn;][]{lee1998}.
Indeed, the presence of presolar
\iso{Be}{10} \citep{mckeegan2000}
suggests that this may well be the case,
since this isotope is in general only produced by cosmic
rays, and not at all by supernovae
\citep[e.g.,][]{vcfo}.
We note that there should also be a steady-state \iso{Be}{10}
abundance in the general ISM.
If cosmic-ray bombardment occurs over
a timescale $\tau_\star \gg \tau_{10} = 2.2 \ {\rm Myr}$,
then immediately afterwards,
the \iso{Be}{10} abundance relative to the stable
isotope is
$\iso{Be}{10}/\iso{Be}{9} \sim \tau_{10}/\tau_\star
  \sim 10^{-3} ({\rm 1\ Gyr}/\tau_\star)$.
To explain the full pre-solar value
$\iso{Be}{10}/\iso{Be}{9} \sim 10^{-3}$
\citep{mckeegan2000}
would seem to require an
effective Galactic cosmic-ray nucleosynthesis
timescale of $\tau_\star \sim 1$ Gyr,
which is short compared to the $\sim 9$ Gyr
cosmic age at the solar birth.
In addition to this Galaxy-wide beryllium production,
the nascent Solar System was likely subject to
transient and localized enhancements in
Galactic (i.e., high-energy) cosmic rays that would lead to some level
of enhancement in
spallation radioisotope production
\citep{fcvn}.
We note that a nearby supernova represents a local
cosmic-ray accelerator.
In addition, as the protosolar core collapses, it becomes
opaque to the already high flux of $\iso{Be}{10}$ Galactic cosmic-rays
and traps them \citep{desch2004}.
Nonetheless, \fe{60} is still the one isotope not produced
at the needed levels in the cosmic-ray scenario,
and thus points to the need for at least {\em some}
supernova material \citep[e.g.,][]{hester2005}.

In any case,
local production (and/or other extra sources, including the possibility
of multiple supernovae)
would mean that the
presolar radioisotope abundance
$X_i$ would have two, or more, components.
In those conditions, the radioactivity distances estimated
in this paper will, by definition, be
lower limits to the supernova distance.
To disentangle these sources is not a trivial
task, but not hopeless either, as we are aided
by (1) the wide range of radioisotope lifetimes,
and (2) the rather different abundance patterns
arising from the various nucleosynthesis mechanisms.
For example, if there is a significant local cosmic-ray
production of, say, \iso{Al}{26}, this would
reveal itself as an anomalously low
radioactivity distance for this isotope relative
to a supernova-only species such as \fe{60}.
Such an analysis is beyond the scope of this paper.

Based on probable secondary sources, we exclude $^{146}$Sm from the
parameter study in \S \ref{chi2}.  The isotope has
the longest half-life in Table \ref{data} by almost
a factor of three.   As suggested by \cite{meyer2000}, meteoritic measurements
of $^{146}$Sm are most likely contaminated by a steady-state component in the ISM
maintained by Type Ia supernovae.  If the abundance is increased,
the derived radioactive distance decreases in Equation \ref{dist}, which explains
its location at the bottom of Figure \ref{rauscher}.  
On the other hand,$^{53}$Mn, although within the
error bars of $^{60}$Fe, demands a high radioactive distance.
Yet, like $^{146}$Sm, the measured $^{53}$Mn abundance is expected to have
been contaminated by the
steady-state ISM abundance \citep{meyer2000}, implying an
even larger distance for the supernova.  
However, $^{53}$Mn is produced
during explosive silicon burning, which occurs at the very end of the
supernova process.  Because the Type-II supernova explosion mechanism
is still poorly understood, it is difficult to accurately predict the
products of explosive silicon burning \citep{woosley2002}.
In addition, the overabundance of $^{53}$Mn in the supernova ejecta has been explained
by the probable fallback of the innermost layers of the star \citep{meyer2005}.
For these reasons, we also exclude $^{53}$Mn from the parameter study in \S \ref{chi2}.

Finally, 
within the context of the model in this paper,
the radioactivity distance estimates
for each species will vary
due to the effect of radioactive decay,
if there is any significant time delay between nucleosynthesis and
incorporation into meteorites.
Of course, for any material injected by a nearby supernova,
{\em some} time must inevitably elapse between
radioisotope production and incorporation into meteorites,
in particular the CAIs.
At minimum, there must be a nonzero time-of-flight between
the supernova and the presolar nebula;
also, the time for creation of small bodies should be
$\sim$ 1 to 10 Myrs based on circumstellar disk
lifetimes \citep[e.g.,][]{hartmann2005}.
If the supernova impact itself initiated the collapse or merely
seeded a nearby cloud of the
presolar nebula \citep{cameron1977}, then there is an additional delay
due to the collapse, which is estimated to take $\sim
0.1$ to 10 Myrs \citep[e.g.,][]{shuppiii,cameron1998,vanhala2000,tassis2004}.

Any delay would imply that the supernova radioisotope signal is
underestimated, and correcting for this effect
decreases the radioactivity distance in Equation~(\ref{dist}).
Thus, the un-delayed results of Figure \ref{rauscher}
represent {\em upper limits} to the distance.
However, adjustment to the radioactive distance
is exponentially sensitive to the isotope lifetime;
the shorter-lived species such as \iso{Ca}{41}
and \iso{Cl}{36}
will vary strongly for modest delays of $\sim 100$ kyr,
while the longer-lived species ($\tau \ga 5$ Myr)
will be essentially unaffected over timescales
appropriate for the full ensemble of isotopes.
For example, this explains why $^{41}$Ca demands the largest
radioactivity distance in Figure \ref{rauscher}; $^{41}$Ca has decayed the most.
On the other hand, because of their short half-lives
the initial \iso{Ca}{41} and \iso{Cl}{36} abundances
are also among the most difficult to determine \citep[e.g.,][]{wadhwa2006}.
Indeed, their estimated abundances in Table \ref{data} are really
lower limits, which implies upper limits in Figure \ref{rauscher}.

\subsection{Parameter Modeling}\label{chi2}

In order to comment on the degree of concordance in
Figure \ref{rauscher}, we define the $\chi^2$ of the data at each
progenitor mass with assumed radioactivity distance 
$(D/R_{\rm SS})_{\rm guess}$
to be 
\begin{equation}
{\chi}^2({M_{\rm SN}}) = 
  \sum_i \frac{\left[ (D/R_{\rm SS})_{i}- (D/R_{\rm SS})_{\rm guess} \right]^2}{\sigma_i^2} 
  \ \ .
\end{equation}
We evaluated a grid of $\chi^2$ values
for $(D/R_{\rm SS})_{\rm guess}$ 
ranging from 1 to 200, in steps of 1.
For each nucleosynthesis model and progenitor
mass, a separate $\chi^2$ value was calculated.  In addition,
a decay time ranging from 0 to 2.5 Myrs, in steps of 0.1 Myrs, was included.
This gives a total of 5200 possible combinations for each progenitor mass.
A parameter set was considered an acceptable fit at a
90\% confidence level, i.e. likelihoods $>$ 10\%.

\begin{figure}
\includegraphics{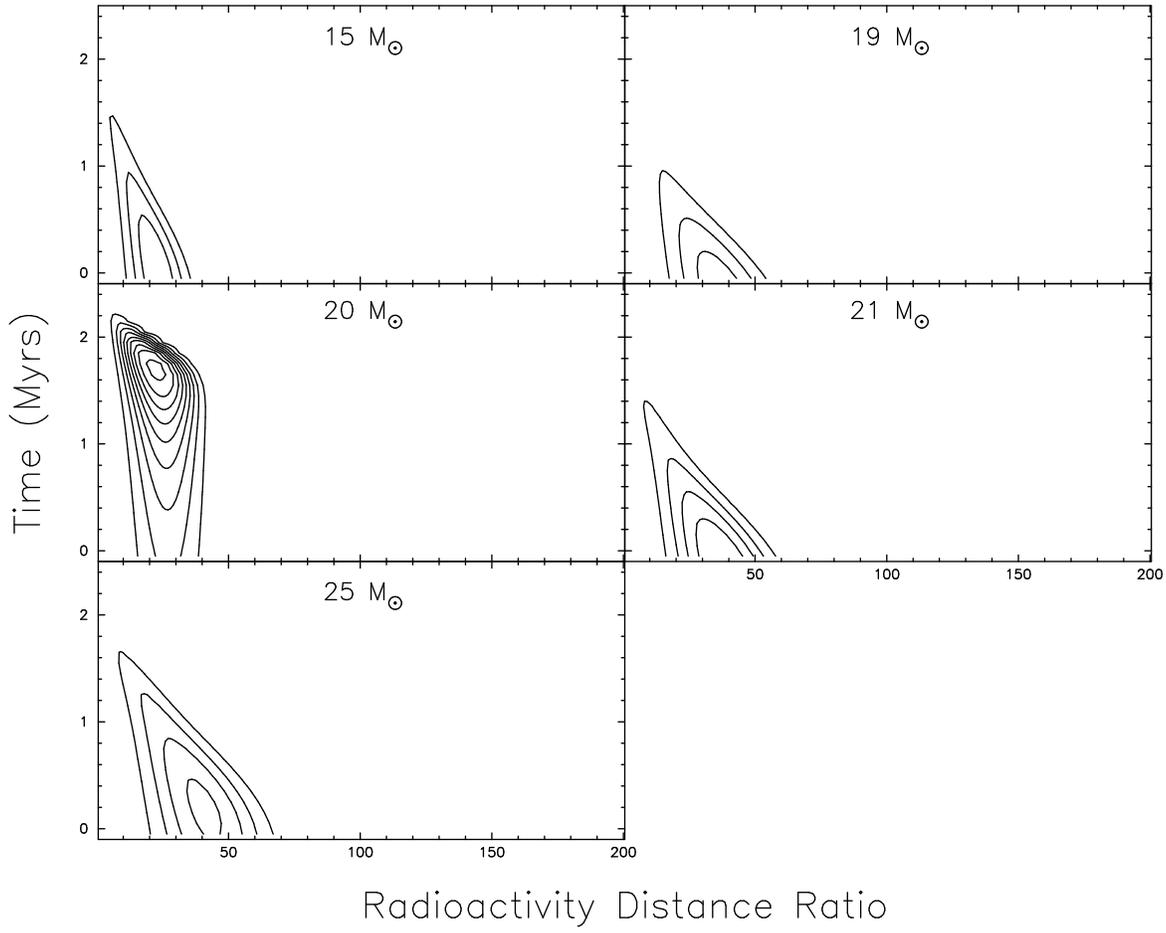}
\vspace{8cm}
\caption{
The likelihoods of the $\chi^2$ parameter search for 5 of the models 
in \cite{rauscher02}.  
The mass of the progenitor in the model is listed in each panel.
The x-axis is the radioactivity distance ratio:
from 1 to 200 in steps of 1.
The y-axis is the time delay from ejection to incorporation into
CAIs: from 0 to 2.5 Myrs in steps of 0.1.
The contours of liklihood are in 10 to 90\% in steps of 10\%.
}
\label{likli}
\end{figure}

Due to the large uncertainty in the data and nucleosynthesis models, there
is a large range of acceptable fits.
The \cite{rauscher02} supernova nucleosynthesis model, with the larger number of
predicted isotopes but smaller mass range, has the largest constraints on
the parameters.  Therefore, we will discuss those fits in detail.
Figure \ref{likli} shows the calculated likelihoods of the parameter
study.  The lowest contours in Figure \ref{likli} are 10\%.

With accepted likelihoods greater than 10\%, the mass is not
well constrained (i.e. all mass models fit); however, it is clear that the mass model
of 20 M$_\odot$ is the highly preferred model.  
In fact, the best fit (likelihood of 93\%) is with a 
progenitor mass of 20 M$_\odot$, decay time of 1.8 Myrs,
and a D/R$_{SS}$ ratio of 22.  However, it is important to keep in mind
that all fits in Figure \ref{likli} are statistically equivalent in our 90\% 
confidence level. 
Although the preferred $\chi^2$ of
the 20 M$_\odot$ model is perhaps
intriguing and should be explored more in detail, it is
probably a conspiracy of terms.
The tangible reason that the 20 M$_\odot$ model fits the data so well
is due to the increased predicted abundance of \iso{Ca}{41} and \iso{Cl}{36}
in that model.  It is crucial to remember that the
strict meteoritic data for \iso{Ca}{41} and \iso{Cl}{36} are upper limits,
which can heavily impact the result by decreasing the decay time necessary
to achieve an overlap. 
Figure \ref{4panel} shows the radioactive distance at four different
decay times, t = 0, 0.1, 1.8, and 2 Myrs, t = 1.8 Myrs was included
as it is the best fit time.  The figure illustrates the evolution with different
decay times. 

It is clear from the above discussion that the most constrained 
aspect of the simplified problem is the radioactivity distance
with some constraint on the decay time.
Using a 90\% confidence level, 
the distance ratio and decay time is constrained to be
\begin{equation}
5 < D/R_{SS} < 66 
\end{equation}
and
\begin{equation}
0 < \tau < 2.2~\rm{Myrs,}
\end{equation}
respectively,
using the
\cite{rauscher02} nucleosynthesis model.  
In order to compare the distance ratio to other distance
estimates, we have to use assumptions for the early solar
system mass and radius.
The large time range is unsurprisingly consistent with many
other estimated time scales
\citep[e.g.,][]{meyer2003,meyer2005,oue2005}.

\section{Implications for the Presolar Environment}
\label{numbers}

The largest assumption in Figure \ref{4panel} is that a supernova
exploded nearby the protosolar nebula such that up to 2.2 Myrs later
those isotopes were incorporated into meteorites. 
The simplest and most probable explanation is that our Sun formed in a loose cluster
containing a massive star that later dispersed.
In that case, was the Sun triggered by the supernova
\citep[e.g.,][]{cameron1977}, or did it form
at a similar time as the progenitor star
\citep[e.g.,][]{chevalier2000}?  

One must keep in mind that the temporal choreography is crucial
to this discussion.  Our results show that the time scale for the isotopical
inclusion into solid objects ranges from 0 to 2.2 Myrs, including
travel, grain growth, and planetesimal formation.  That is a
particularly quick timescale when compared to the lifetime
of a protostar.
In addition, if we consider that the expected timescale for large body
formation occurs at the end of the protostellar
lifetime, during the Class II/III stage or later 
\citep[e.g.,][]{hartmann2005},
then the likelihood of
this particular supernova triggering our solar system does exist, but
it must evolve quickly.
Nonetheless, we discuss these scenarios using the current best estimates
of early nebula radius and mass.

\subsection{Siblings at Birth}

In the case of siblings, the protosun and the progenitor star are 
formed on similar time scales \citep[within $\sim$ 1~Myrs,][]{ladalada2003}: 
possibly a birth case similar
to the cluster of about 1600 low-mass
young stars in the Orion Nebula within $\sim$2 pc of the
the Trapezium group \citep[core radius of 0.2 pc,][]{hillenbrand1998}.
Within the inner 0.4 pc, 85\% of the stars have L-band excesses that
suggest circumstellar disks \citep{lada2000}.
In spite of the harsh environment near the massive cluster, the
low-mass young stars near the central star
$\theta^1$ Ori C have observable disks due to the
photoionization effects of the massive star \cite[e.g.,][]{mccaughrean1996}.

\begin{figure}
\includegraphics{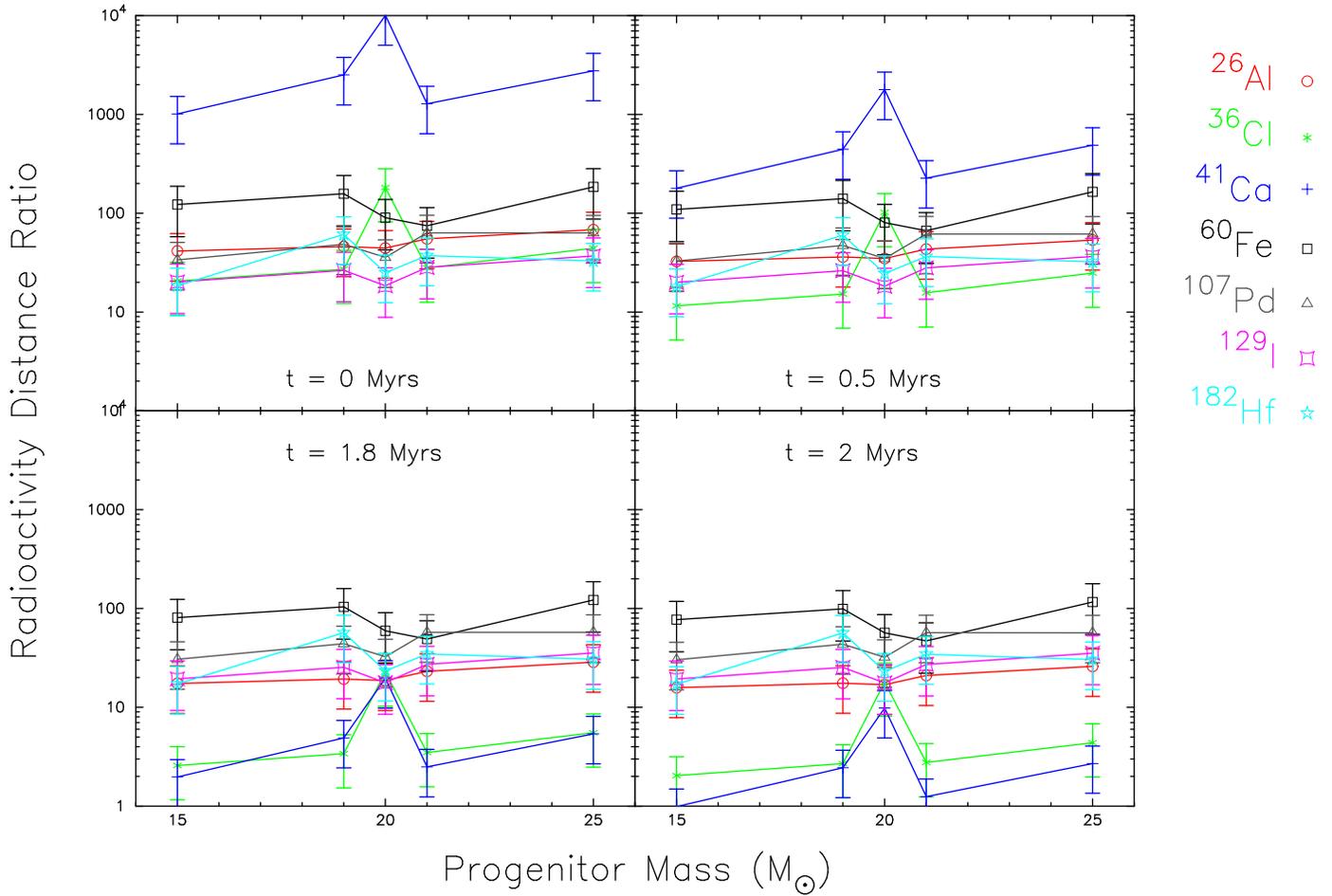}
\vspace{8cm}
\caption{
The Radioactive Distance ratio with a decay time of 0, 0.1, 1.8, and 2 Myrs
for the \cite{rauscher02} radionucleosynthesis model.
Note that $^{146}$Sm and $^{53}$Mn have been excluded for reasons described in the
text.
}
\label{4panel}
\end{figure}

Clearly from Figures \ref{likli} and \ref{4panel}, the isotopic
evidence does not well constrain the mass of the progenitor star.  
Nonetheless, the main sequence lifetime of the progenitor star can be estimated 
from recent models \citep[e.g.,][]{romano2005} as 7 to 10 Myrs for   
the mass range in Figure \ref{4panel}. 
In addition to the main sequence lifetime, one must include the protostellar lifetime of
the massive star.  Although this value is poorly known, the best estimates
range from $10^4$ to $10^5$ years \citep[e.g.,][]{yorke1986,mckee2002},
a fraction of the main sequence lifetime.

Although the time frame of $\sim$10 Myrs is significant,  
low-mass stars evolve at a much slower pace than higher mass stars.
In fact, the longest stage may be before collapse even begins
on the order of the ambipolar diffusion
timescale, which ranges from 0.1 to 10 Myrs, depending on
the initial mass-to-flux ratio and degree of ionization in the cloud
\citep[e.g.,][]{tassis2004}.  That in combination with the standard 
spread of ages in clusters of $\sim$1 Myrs \citep{ladalada2003} 
implies that the protostellar
nebula could have been in many possible evolutionary stages
\citep[also see][]{hester2005}:
(1) the youngest protostellar evolution stage, 
the so-called
Class 0 stage, which is dominated by an circumstellar envelope that implies
$R_{\rm SS} \simeq $ 5000 AU and $M_{SS} \simeq $ 1~M$_\odot$ \citep[e.g.,][]{looney2000,looney2003}.
Although the injection efficiency is probably closer to 10\% \citep{vanhala2002}, 
we conservatively use $\finj \simeq 1$ (i.e., larger distance); or
(2) an older protostar, Class I/II, 
dominated 
by the circumstellar disk that implies $R_{\rm SS} \simeq $ 100 AU and $M_{SS} \simeq$ 0.02~M$_\odot$
\citep[e.g.,][]{looney2000}, which is similar to the
minimum mass solar circumstellar disk \citep[e.g.,][]{weidenschilling1977}.  
We conservatively assume a large injection efficency
of $\finj \simeq 1$. 

From our analysis in \S \ref{chi2}, the radioactivity distance ranges from 5 to 66.  Using that
range of radioactivity distance ratio fits with the above size and mass estimates,
yields distances of 0.12 to 1.6 pc and 0.02 to 0.22 pc, respectively
\citep[cf.][]{chevalier2000,oue2005}.
Note that a very low disk mass in older Class III sources or even debris disks 
would result in a supernova distance that is many pc away in this formalism
(i.e. M$_{SS}$ is very small).
However, it is unclear how those systems could
manufacture the meteorites necessary for the short lived isotope detection.
The ejecta must be incorporated into the objects at an early enough
time to be detected in meteorites.
It is also important to note that these derived distances are very consistent with
low-mass systems seen in clusters with large mass members (e.g., Orion or Eagle Nebula).

\subsection{Triggered Birth}

If we assume that a supernova event injected short lived isotopes into
the early solar nebula, one must also consider that this event could have
triggered the collapse of the nebula \citep[e.g.,][]{cameron1977}.
Many recent models have shown that moderately slow shock waves
(20-45 km/s) can trigger the collapse of cores
\citep[e.g.,][]{boss1995,cameron1998}: too little and no collapse, too much and
the core is torn apart.
A quick collapse process \citep[$\sim 10^4 - 10^5$ yrs,][]{cameron1998}
may also account for the low radioactive distance (i.e. large
$X_{i}$) of the
majority of the longer half-life isotopes 
($^{146}$Sm, $^{129}$I, $^{182}$Hf, and $^{107}$Pd)
seen in Figure \ref{rauscher}; the molecular
cloud core in which the Sun formed would not have time to decay the
initial ISM component.  
On the other hand, \citet{hester2005}
argue that a supernova triggering a collapse and injecting  that
collapse with its radioisotopes, 
is probably less common than the scenario where
supernova ejecta intercept a protostar. 

In the triggering scenario, 
the solar nebula would not yet have evolved to 
a Class 0 protostar, implying two possible states:
(1) a starless core,
which has $R_{\rm SS} \simeq$ 7000 AU and typical mass
of $M_{SS} \simeq $ 2~M$_\odot$ \citep[e.g.,][]{jason}.
Again, although the injection efficency is probably closer to 10\% \citep{vanhala2002},
we conservatively use $\finj \simeq 1$ (i.e., larger distance); or
(2) a quickly triggered (e.g. $<$ 1 Myrs), diffuse cloud as modeled by
a Bonner-Ebert sphere, considered
by \cite{vanhala2000,vanhala2002},
with $R_{\rm SS} \simeq$ 0.06 pc, mass of $M_{\rm SS} \simeq 1 M_\odot$, and $\finj \simeq$ 0.1.
With these two possibilities, the estimated ranges of distances are 0.12 to 1.6 pc and
0.06 to 1.2 pc (or 0.3 to 4 pc for the overly conservative case of $\finj \simeq 1$), respectively.  
This can be compared
to the values of \cite{vanhala2000,vanhala2002}, 5.2 to 26 pc.  
\cite{vanhala2000,vanhala2002} comment on the mismatch of distance and
offer possible resolutions.
On the other hand, the possibility of a supernova inducing
a quick collapse of a very diffuse cloud, i.e., $R_{SS} >$ 0.1 pc, will
significantly increase the derived distance to the supernova.

\subsection{Travel Time}

With an assumed ejecta speed of 30 km/s and the maximum
estimated distances from above (e.g., 1.6 pc or even most conservatively 4 pc), the
travel time can be estimated at $\sim 5 \times 10^4$ and
$1 \times 10^5$ yrs, respectively.  Using our best fit
decay time of 1.8 Myrs, this implies that the ejecta material mixed with the
solar nebula material until meteorites were produced.
Given nearly 1.8 Myrs of residency
within the solar nebula, it is possible that
the SN ejecta would become homogenized in the cloud.
This would be consistent with the formation timescale of
calcium-aluminum-rich inclusions (CAIs), on the order of a million years
\citep[e.g.,][]{wadhwa2000}.

\section{Conclusions}

We present a simple model relating the pre-solar abundances
of short-lived radioisotopes to the 
properties of the injecting supernova event. 
While our model is idealized, it is also explicit and computationally
simple, yielding a parameter study whose
general aspects are intriguing.  
In general, we are driven to conclude that a very 
nearby supernova event must be invoked
if {\em any} presolar radioactivities are due to 
injection from a single explosion.
Specifically, 
to a formal confidence level of 90\%, the distance from
the Solar Nebula and the supernova was in the range of 
0.1 to 1.6 pc, for siblings and up to approximately 4 pc (in the most
conservative, and unlikely, case of perfect injection) for triggered birth,
using the \cite{rauscher02} nucleosynthesis
models for 15 to 25 M$_\odot$ progenitors.
This sounds surprisingly close, but
it is consistent with typical
distances found for low-mass stars clustering around one or more
massive stars.
We posit that our Sun was a member of such a cluster that has
since dispersed.

It is important to note that our distance estimates imply that the Sun was directly
influenced by the massive star's photoionization effects, much like 
proplyds in Orion.  To understand the Sun's formation, one must
include the large role played by such a complicated environment.
The minimum distance is also interesting, as it is similar
to the distance at which the disk would probably be destroyed $<$ 0.25 pc
\citep{chevalier2000}.
Our estimated distance range includes the expected uncertainties of the
nucleosynthesis models and the radioisotopic measurements.
Although the physical parameters of the solar nebula also play
a large role in these estimates, we use nearly all
protostellar evolutionary states to compile our range.

In addition to the distance of the supernova, we constrain
the time scale of the explosion to the creation
of small bodies to a 90\% confidence range of
0 to 2.2 Myrs.
The majority of this time was probably spent in our solar system
during the building of small objects from the circumstellar dust;
in fact, it is consistent with the estimates for chondrule production.
It is important to note that the temporal choreography from supernova
ejecta to meteorites is crucial.
The upper limits of our fit decay times make it possible for the
supernova to have triggered the fast collapse of our Sun,
seeded it with short-lived isotopes, and created CAIs,
but it seems more
likely that the Sun was already a protostar during
the explosion \citep[e.g.,][]{hester2005}.

In addition to the inherent uncertainties in our model,
there are additional assumptions that would imply a smaller or
larger supernova distance.
There are two clear cases that would cause our deduced distances
to be an upper limit:
(1) an inclined circumstellar disk (Class I/II source) would have a lower
cross-sectional area and not collect the full 
supernova contaminates as a face-on disk, and (2)
the ejecta may not be effectively
stopped by the circumstellar disk as assumed ($\finj \ll 1$)
\citep[i.e., effectiveness of the injection process; see][]{vanhala2002}.
Similarly, there are two opposing cases that would make our
deduced distances lower limits:
(1) highly inhomogeneous ejecta 
\citep[i.e. bullets as seen in Cas A,][]{willingale2003}
would not allow a predictable distance,
and (2) although the material is injected uniformly onto
the circumstellar disk, the inner regions of the disk has 
a higher
density than the outer regions \citep[e.g.,][]{mundy1996}, so if there
is no mixing, the measured meteorites are produced {\it in situ}
with abundances lower than expected.
These crucial issues lie within the domain of sophisticated
(magneto)hydrodynamical simulations that will yield
a fuller understanding of this problem;
our simple model serves as a tool that
highlights and quantifies the issues that the 
full numerical models address.

In closing, we note that the analysis presented here
for extinct presolar meteoritic radioactivities
closely parallels considerations of live
geological radioactivities
as signatures of relatively recent nearby supernovae
\citep[e.g.,][]{efs,knie}.
Both analyses leverage the uniqueness of radioisotopes as
a signature of supernova activity, and
both use the inverse square law in the same way to gauge distance.
Moreover, 
\citet{fhe} emphasize that
given a confirmed nearby supernova event,
one can then 
turn the problem around,
reading the presence or absence of, and detailed abundances of,
all radioisotopes as direct probes of supernova ejecta and
thus nucleosynthesis.
This adds another way that meteoritic anomalies 
serve as fossil testimony to our
massive sibling.

\acknowledgements{
We thank You-Hua Chu, Charles Gammie, and Konstantinos Tassis for discussions.
In particular, we thank the referee for significantly impacting the quality
and usefulness of this paper.
L.W.L. acknowledges support from the Laboratory for
Astronomical Imaging at the University of Illinois,  
the National Science Foundation under Grant No. AST-0228953,
and NASA under Origins grant No. NN06GE41G.
B.D.F. acknowledges support by the National Science
Foundation under Grant No. AST-0092939.}

\bibliographystyle{apj}
\bibliography{supernova}
\clearpage

\begin{deluxetable}{llcrrllll}
\tabletypesize{\scriptsize} 
\tablewidth{0pt}
\tablecaption{Adopted Short-lived Radioactive Isotopes\label{data}}
\tablehead{
  \colhead{Radio-\tablenotemark{a}} & \colhead{Ref.} & \colhead{$t_{1/2}$} &
  \colhead{Meteoritic\tablenotemark{b}} & \colhead{Meteoritic\tablenotemark{b}} & 
  \colhead{Reference\tablenotemark{c}} & \colhead{Reference\tablenotemark{c}} &
  \colhead{Mass} & \colhead{Mass}\\
  \colhead{isotope}  & \colhead{} & \colhead{} & \colhead{Ratio} &
  \colhead{Uncert.} & \colhead{Abundance} & \colhead{Uncert.} & \colhead{Fraction} &
  \colhead{Uncert.} \\
  \colhead{$^i{\cal P}$} & \colhead{${\cal P_{\rm ref}}$} & \colhead{(Myr)} &
  \colhead{$^i{\cal P/P_{\rm ref}}$} & \colhead{$\sigma_{^i{\cal P}}$} &
  \colhead{${\cal P_{\rm ref}}/{\rm H}$} & \colhead{$\sigma_{P_{\rm ref}}$} & 
  \colhead{$X_i$} & \colhead{$\sigma_{X_i}$}}
\startdata
$^{26}$Al & $^{27}$Al & 0.72 & $5.9 \times 10^{-5}$& 
    $0.6 \times 10^{-5}$& $3.46 \times 10^{-6}$ & $0.16 \times 10^{-6}$ &
    $3.77 \times 10^{-9}$ & $0.42 \times 10^{-9}$\\
$^{36}$Cl & $^{35}$Cl & 0.30 & $1.6 \times 10^{-4}$ & 
    $0.7 \times 10^{-4}$& $2.15 \times 10^{-7}$ & $0.29 \times 10^{-7}$ &
    $8.82 \times 10^{-10}$ & $4.04 \times 10^{-10}$\\
$^{41}$Ca & $^{40}$Ca & 0.10 & $1.4 \times 10^{-8}$ & 
    $0.1 \times 10^{-8}$& $2.59 \times 10^{-6}$ & $0.18 \times 10^{-6}$ &
    $1.06 \times 10^{-12}$ & $0.10 \times 10^{-12}$\\
$^{53}$Mn & $^{55}$Mn & 3.7 & $2.8 \times 10^{-5}$ & 
    $0.3 \times 10^{-5}$& $3.77 \times 10^{-7}$ & $0.26 \times 10^{-7}$ &
    $3.98 \times 10^{-10}$ & $0.51 \times 10^{-10}$ \\
$^{60}$Fe & $^{56}$Fe & 1.5 & $7.5 \times 10^{-7}$ & 
   $2.5 \times 10^{-7}$ & $3.45 \times 10^{-5}$ & $0.24 \times 10^{-5}$ &
   $1.10  \times 10^{-9}$ & $0.38 \times 10^{-9}$ \\
$^{107}$Pd & $^{108}$Pd & 6.5 & $2.0 \times 10^{-5}$ & 
   $0.2 \times 10^{-5}$ & $5.90 \times 10^{-11}$ & $0.41 \times 10^{-11}$ &
   $8.98 \times 10^{-14}$ & $1.09 \times 10^{-14}$ \\
$^{129}$I & $^{127}$I & 15.7 & $1.1 \times 10^{-4}$ &  
   $0.1 \times 10^{-4}$ & $4.10 \times 10^{-11}$ & $1.13 \times 10^{-11}$ &
   $4.14 \times 10^{-13}$ & $1.20 \times 10^{-13}$ \\
$^{146}$Sm & $^{144}$Sm & 103 & $7 \times 10^{-3}$ & 
   $2 \times 10^{-3}$ & $1.05 \times 10^{-11}$ & $0.10 \times 10^{-11}$ &
   $7.60 \times 10^{-12}$ & $2.28 \times 10^{-12}$\\
$^{182}$Hf & $^{180}$Hf & 8.9 & $1.1 \times 10^{-4}$ & 
   $0.1 \times 10^{-4}$ & $6.99 \times 10^{-12}$ & $0.64 \times 10^{-12}$ &
   $9.95 \times 10^{-14}$ & $1.28 \times 10^{-14}$ 
\enddata
\tablenotetext{a}{Only those isotopes suspected to come from core-collapse supernovae \citep[e.g.,][]{wadhwa2006}
are included, except $^{92}$Nb and $^{244}$Pu.
$^{92}$Nb and $^{244}$Pu were omitted because of a very high uncertainty
and a lack of predictions in the two supernovae ejecta models used in this paper, respectively.
The ratio given here is the initial solar system abundance, which
is measured at the creation of the CAIs, not necessarily at isotopic closure.
}

\tablenotetext{b}{Other references used are
$^{26}$Al: \cite{young2005,young2005b,cos2006};
$^{36}$Cl: \cite{lin2005} (The ratio used is 
dependent on the $^{26}$Al/$^{27}$Al ratio in the measured mantle compared to the
CAI. The ratio used was consistent, within the error bars, with our assumed
value.  However, the $^{26}$Al/$^{27}$Al ratio in the mantle was not well detected, so this
value is more correctly taken as a lower limit.);
$^{41}$Ca: \cite{srinivasan1996}  (This ratio was measured in a CAI with a measured
$^{26}$Al/$^{27}$Al ratio that was consistent, within the error bars, 
with our assumed value.  However, it is
important to note that if that particular CAI formed late, even by 
a small amount, the measured $^{41}$Ca abundance ratio would be heavily impacted.
Again, this ratio is more correctly taken as a lower limit.);
$^{53}$Mn: \cite{nyquist1999};
$^{60}$Fe: \cite{tachibana2006};
$^{107}$Pd: \cite{hauri2000,carlson2001} (There is a suggestion that the initial
abundance ratio may have been as high as $40 \times 10^{-5}$.);
$^{129}$I: \cite{brazzle1999};
$^{146}$Sm: \cite{lugmair1992};
$^{182}$Hf: \cite{kleine2005,kleine2006}.
Often the published error bars are 2$\sigma$, but to be conservative
we used the errors as 1$\sigma$.  If the estimated error was less than 10\%, we
used a conservative 10\% error.
}

\tablenotetext{c}{From Table 2 \cite{lodders2003}, using the initial solar system abundances, i.e.
correction for gravitational settling.}
\end{deluxetable}

\clearpage

\clearpage

\clearpage

\clearpage

\end{document}